\documentclass[%
 reprint,
 amsmath,amssymb,
 aps,
]{revtex4-2}

\usepackage{graphicx}
\usepackage{dcolumn}
\usepackage{bm}

\usepackage{graphicx}
\usepackage{grffile}
\usepackage{dcolumn}
\usepackage{bm}
\usepackage{siunitx}
\usepackage{hyperref}
\usepackage{color}
\usepackage{multirow}

\newcommand{\beq}{\begin{equation}}
\newcommand{\eeq}{\end{equation}}

\makeatletter
\def\@fnsymbol#1{\ensuremath{\ifcase#1\or \dagger\or *\or \ddagger\or
   \mathsection\or \mathparagraph\or \|\or **\or \dagger\dagger
   \or \ddagger\ddagger \else\@ctrerr\fi}}
    \makeatother

\begin{document}

\preprint{APS/123-QED}

\title{
Molecular insights into the physics of poly(amidoamine)-dendrimer-based supercapacitors}
\author{Tarun Maity} 
\affiliation{Center for Condensed Matter Theory, Dept. of Physics,\\ Indian Institute of Science, Bangalore 560012, India\\}
\author{Mounika Gosika}
\altaffiliation{Center for Condensed Matter Theory, Dept. of Physics,\\ Indian Institute of Science, Bangalore 560012, India}
\affiliation{Centro de F\'{\i}sica de Materiales (CSIC, UPV/EHU) and Materials Physics Center MPC, Paseo Manuel de Lardizabal 5, 20018 San Sebasti\'{a}n, Spain\\}

\author{Tod A. Pascal}
\affiliation{ATLAS Materials Physics Laboratory, Department of NanoEngineering and Chemical Engineering,\\ University of California, San Diego, CA 92023\\ }
\author{Prabal K. Maiti}
\email{maiti@iisc.ac.in}
\homepage{http://www.physics.iisc.ernet.in/~maiti/}
\affiliation{
Centre for Condensed Matter Theory, Department of Physics, Indian Institute of Science, Bangalore 560012, India
}%

\date{\today}

\begin{abstract}
Increasing the energy density in electric double-layer capacitors (EDLCs), also known as supercapacitors, remains an active area of research. Specifically, there is a need to design and discover electrode and electrolyte materials with enhanced electrochemical storage capacity. Here, using fully atomistic molecular dynamics (MD) simulations, we investigate the performance of hyper-branched ‘poly(amidoamine) (PAMAM)’ dendrimer as an electrolyte and an electrode coating material in a graphene-based supercapacitor. We investigate the performance of the capacitor using two different modeling approaches, namely the constant charge method (CCM) and the constant potential method (CPM). These simulations facilitated the direct calculation of the charge density, electrostatic potential and field, and hence the differential capacitance. We found that the presence of the dendrimer in the electrodes and the electrolyte increased the capacitance by about $65.25 \%$ and $99.15\%$ respectively, compared to the bare graphene electrode-based aqueous EDLCs. Further analysis revealed that these increases were due to the enhanced electrostatic screening and reorganization of the double layer structure of the dendrimer based electrolyte. 
\end{abstract}
\maketitle

\section{\label{sec:level1} Introduction}
    
Improving the energy densities of electric double layer capacitor (EDLC), also known as supercapacitor based energy devices, is of great practical interest ~\cite{vatamanu2015capacitive,vatamanu2015non,vatamanu2011influence}, as these devices have higher power densities compared to conventional energy storage devices such as fuel cells, electrochemical batteries and even dry/moist electrolytic capacitors. The improved energy density, in turn, would increase the commercial viability of EDLCs, thus spurring large scale, mainstream adoption ~\cite{conway2013electrochemical}.
    To this end, recent work has focused on engineering advanced electrode and electrolytes materials.
Porous materials, with high specific surface areas and good electronic conductivity, have traditionally been used as electrode materials in these devices ~\cite{barpanda2011structure,zhi2013nanostructured}. Room temperature ionic liquids (RTILs) are typically chosen as the electrolyte, owing to their superior electrochemical properties, including high  operating voltage windows and non-flammability, compared to aqueous and organic based electrolytes ~\cite{simon2010materials}. Yet, despite progress over the years in electrode and electrolyte optimization, further advances demand direct knowledge of the interfacial structure and dynamics, and design principles for unique nanoscale physics therein.

    In this work, we consider dendrimers: hyper-branched synthetic polymers as intriguing potential candidates as electrolytes in these devices ~\cite{li2000visualization}. Dendrimers have well-defined molecular structure, are flexible in size and shape, and are responsive to controllable stimuli such as pH ~\cite{esfand2001poly, maiti2004structure, maiti2005effect}. They are also known to undergo structural deformations at interfaces ~\cite{lin2015nitrogen,gosika2018ph} and their charge densities are comparable to that of ionic liquids ~\cite{lin2013poly}. Moreover, the pore size of a typical microporous electrode ranges from 0.5 to 5 nm, which matches well with the sizes of poly amidoamine (PAMAM) dendrimers, which vary from 0.6 nm for generation 0 (G0) to 6 nm (G10) ~\cite{gosika2018ph, maiti2005effect, maingi2012dendrimer,maiti2004structure}. Experiments and theory have shown that the capacitance in EDLCs is significantly enhanced when the pore size of the electrode material matches the size of the electrolyte ions ~\cite{largeot2008relation, merlet2012molecular}. 
    
    Interfacial adsorption of dendrimers on electrodes dynamically exposes their charge groups to the electrolyte, which facilitates the formation of unique electric double layer structures. This has been demonstrated by experiments by Guo \textit{et al.} ~\cite{guo2006high}, who reported that hyper-branched polymers, like dendrimers, exhibit very low losses in the dielectric response function, even at high operational frequencies ($\sim$ 1 MHz). Freire  \textit{et al.} also showed that the presence of dendrimer can screen repulsive contacts between different counter ion  molecules and favored ionic conductivity ~\cite{freire2013molecular}.
    These interesting properties of dendrimers make them potentially excellent candidates as electrolytes in EDLCs, however, a thorough microscopic examination of the interfacial behaviour, and the resulting effect on electrochemical performance, have not yet been elaborated.  
    
    Computer simulations employing molecular dynamics (MD) based quantum mechanically derived accurate potentials (force-fields) is a common tool for elucidating the microscopic nature of interfacial systems, and would be well suited for exploring the role of dendrimers in modulating the EDL structure and ultimately the performance of EDLCs. The key here is an accurate description of the interaction parameters, coarse-grained simulations revealed significant force-field dependence in the binding strength of dendrimers to graphene electrodes, as the system’s pH ~\cite{maiti2005effect} is varied ~\cite{maerzke2015deformation}. To more clearly understand the nature of the interaction of the dendrimer-graphene composites requires us to go beyond coarse-grained force-fields and perform fully atomistic simulations. Yet, fully atomistic simulations of dendrimers at interfaces, where the dendrimer is being used as an electrolyte, are relatively rare to the best of our knowledge. In contrast, there have been several experimental~\cite{trigueiro2016carbon,rath2015reduced} and simulation studies~\cite{shim2011graphene,salanne2016efficient} that reported carbon-based electrode materials and studied the capacitance values with ionic liquids being the electrolytes ~\cite{mao2019self}.
    For example, Triguerio \textit{et al.} ~\cite{trigueiro2016carbon} reported that the dendrimer functionalised carbon nanotubes can improve the nanotube’s performance as an electrode. Another study by Chandra \textit{et al.} ~\cite{chandra2015dendrimer} reported that dendrimer functionalised nanoparticles coated on an electrode surface can enhance the surface area available to the electrolyte atoms, thereby achieving efficient charge transfer and low contact resistances.
    In another experimental work, Liu \textit{et al.} ~\cite{liu2017ultrathin} used dendrimer functionalized graphene-oxide sheet as
    a coating on the sulfur electrode of a Li-S battery and achieved long cycle life (up to 500 cycles). When considering common electrolytes in EDLCs, various computational studies on RTILs have been reported, including work by Yeh \textit{et al.} ~\cite{yeh1999ewald} which discussed the effect of periodic boundary conditions (BCs) in EDLC simulations. 
    
    Recently, we used atomistic MD simulations to study the structural deformations~\cite{gosika2018ph} and the free-energy of the binding ~\cite{gosika2019understanding} of PAMAM dendrimers at a charge neutral graphene/water interface, as a function of the protonation state of the dendrimer. We found that the van der Waals interactions play a pivotal role in driving dendrimer adsorption. We also found that moderately charged (neutral pH) dendrimers achieve maximum surface wetting as compared to the non-protonated (high pH) and fully protonated (low pH) dendrimers. We showed that lower generation dendrimers tended to deform and form flat, disk-like architectures, with good surface accessibility, at the graphene/water interface ~\cite{gosika2018ph}. These observations suggest that the lower generation PAMAM at neutral pH condition as an ideal choice for achieving maximum charge densities in PAMAM-based supercapacitors. Therefore, in this work, we considered a G2 PAMAM dendrimer at neutral pH, and elucidate the electrochemical performance in graphene/water based EDLCs.   
    Beyond accurate simulations of dendrimer based systems, we are also concerned with modelling biased nanoscale interfaces, as a means of probing electrochemical effects in EDLCs. Here, there are two main computation methods commonly employed for doing this: 1) the constant charge method (CCM) – the charges of the electrode atoms are fixed, and 2) the constant potential method (CPM) – a grand canonical statistical mechanical ensemble is defined  by means of a fictitious bath that exchanges electrons with the electrodes to maintain a constant electrode potential (the number of electrons and chemical potential are conjugate pairs) ~\cite{wang2014evaluation,merlet2013computer}. The CPM approach is generally preferred as it  enables simulations that are more directly comparable to experiments. However, it is somewhat restricted in its applicability due to significant additional computational demands. To this end, Wang \textit{et al.} ~\cite{wang2014evaluation} compared both approaches for a LiClO\textsubscript{4}-acetonitrile/graphite EDLC and showed that both the approaches lead to similar ion and solvent density profiles for voltages less than $2 V$. Comparing the performance of both approaches for a more complicated electrode/electrolyte morphology is one of the aims of this study. Moreover, to address the computational challenges, Reed and coworkers\textit{et al.} ~\cite{reed2007electrochemical} have developed an efficient approach for simulating cells within CPM~\cite{wang2014evaluation} in the LAMMPS simulation engine, which we employ here.
    
    The manuscript is organized as follows. In section II, we provide the model building and the simulation methodologies adopted in this work. In section III, we present our results on the electrostatic potential, charge density profiles and the capacitance values, obtained from the CCM and the CPM approaches. Finally, in section IV, we summarize our findings and conclude with the key insights from our study to provide future research directions. 
    
    \begin{table}[h]
    \centering
    
    \caption{Details of the electrode-electrolyte combinations studied. The structure of the dendrimer considered corresponds to a G2 PAMAM at neutral pH in all cases.}
     \begin{ruledtabular} 
    \begin{tabular}{ c  c  c  c  }

            \centering
             System  & Positive & Negative  & Electrolyte\\ 
                     & Electrode($+15 e$) & Electrode($-15 e$) &            \\ [2.5ex]
                     
             \hline        

             BAREGP  & Graphene  & Graphene  & H\textsubscript{2}O \\ [1.5ex]

             G2PEL & Graphene  & Graphene &    H\textsubscript{2}O + \\
             &  &  & G2 PAMAM \\ [0.5ex]
             &  &  &  + Cl\textsuperscript{-} ions \\ [0.5ex]

             GPDEN & Dendrimer coated & Graphene & H\textsubscript{2}O \\
                   &  graphene & &  \\ [0.5ex]
        \end{tabular}
        \label{tab:terminology}
         \end{ruledtabular} 
    \end{table}


\section{\label{sec:level2} Model and Simulation Details}
  In this work we focus on an aqueous supercapacitor simulation cell based on 
    water/water+dendrimer electrolyte and graphene/dendrimer coated graphene electrode (Table~\ref{tab:terminology}). 
    The initial structure of the PAMAM dendrimer was built using the Dendrimer Builder Toolkit (DBT)~\cite{maingi2012dendrimer}, while VMD~\cite{humphrey1996vmd} was used to build the graphene sheet. The aromatic carbon atoms of graphene were described using the AMBER FF10 force field (atom-type CA), as in our earlier works ~\cite{kumar2018phase,majumdar2021dielectric}, which we showed captures the water-graphene interactions accurately. 
    Inter-molecular interactions involving dendrimer atoms were described by GAFF force field~\cite{wang2004development}, the 
    water molecules were modelled using TIP3P~\cite{jorgensen1983comparison, price2004modified} water model, and the Joung-Cheatham~\cite{joung2008determination} parameters were used to model the Cl\textsuperscript{-} counter ions.
     Using the xLEaP module of AMBER 14 ~\cite{case2014amber}, the equilibrated electrolyte solution was placed in-between two graphene sheets with dimensions of 32.63 \AA $  \times  $ 62.50 \AA \, and separated by 150 \AA.
 
   \begin{figure}[h!]
   \begin{center}
        \includegraphics[width=\linewidth]{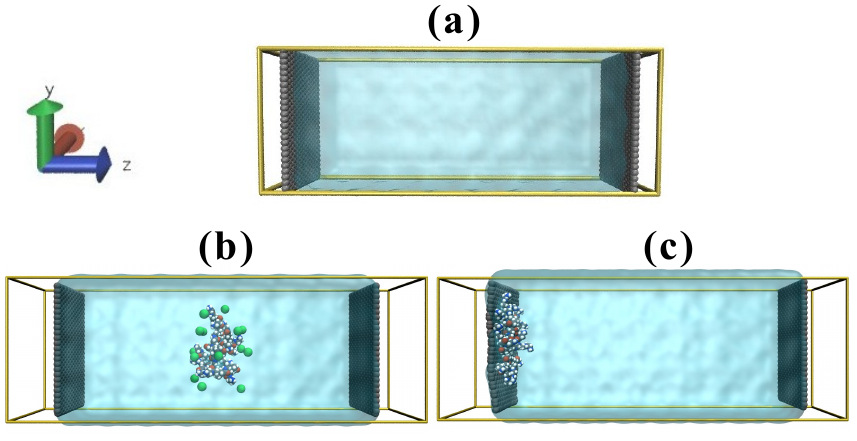}
        \caption{
        Instantaneous snapshot of a simulation box representing (a) BAREGP (b) G2PEL (c) GPDEN at 0 ns: positive electrode is on the left and negative is on the right. The dendrimer is shown in CPK.  The graphene sheet is shown in grey. The protonated amines of the dendrimer are shown in blue and the Cl\textsuperscript{-} counter ions are coloured green.  \SI{15}{\nano\meter} vacuum slab shown, are not drawn to scale.  }

   \label{system}
   \end{center}
    \end{figure}
    
      We aim to understand how the presence of the dendrimer modifies the electrochemical charge storage and the capacitance values of the aqueous supercapacitor with a graphene electrode. Hence, we performed the simulations for a system with a two bare graphene electrode encapsulating a box of TIP3P waters, denoted BAREGP, as a control. To test the potential of using the dendrimer as an electrolyte and as an electrode, two additional systems, namely G2PEL and GPDEN respectively, were considered. G2PEL comprised a G2 PAMAM dendrimer at neutral pH, immersed in a pre-equilibrated box of 10,118 TIP3P water molecules and placed between two graphene electrodes. We neutralize the system by adding  16  Cl\textsuperscript{-} counter ions (the charge of the G2 PAMAM at neutral pH is $+16 e$ )~\cite{maiti2004structure}.
     For the GPDEN system, we covalently grafted the G2 dendrimer onto the graphene electrode, using a "top-binding and grafting-to" approach, as detailed  in our previous work ~\cite{gosika2020covalent}.
     Here, a dendrimer grafted graphene was used as a positive electrode while a pristine graphene was used as a negative electrode.  $10,775$ TIP3P water molecules were added in between the two electrodes.

   \subsection{Constant Charge Method (CCM)}
   
  We employed two different computational schemes for simulating applied bias in our systems. First, we used the PMEMD module of the AMBER14~\cite{case2014amber} MD simulation suite to perform supercapacitor simulations with the CCM method. As mentioned above, in the GPDEN case, the net  charge of the dendrimer coated graphene electrode was $+15 e$, where charge of grafted dendrimer atom was adapted from Gosika \textit{et al.}~\cite{gosika2020covalent}. 
  Thus, a charge of $ 0.01875 e $ was distributed on each of the carbon atom comprising the graphene electrode. In these studies, we minimized spurious electrode - electrode electrostatic interactions by inserting large 15 nm vacuum slab buffer in $z$-direction between the simulation cell, as shown in Fig.  \ref{system}. For the BAREGP and G2PEL cases, we performed two sets of simulations, where the electrodes are 
   i) discharged (charge on every electrode atom is set to $0 e$) ii) oppositely charged (charge on each electrode atom was set to $0.01875 e $, to be consistent with the GPDEN system ). 
    We initiated each simulation with 1000 steps of the steepest descent energy minimization, followed by further 1000 steps of conjugate gradient minimization, to remove the initial bad contacts. We then gradually heated the system from 0 K to 300 K using a Langevin thermostat with random friction collision frequency \SI{1}{\per\pico\second} in the constant temperature-constant volume (canonical or NVT) ensemble for 5 ns. During heating, we restrained the solute (electrode) atoms with a harmonic spring, with a force constant of  20 kcal mol\textsuperscript{-1} \AA\,\textsuperscript{-2} 
    Equilibrium MD simulations  were then performed under the constant pressure-constant temperature (isothermal-isobaric or NPT) ensemble, with a weaker restraint (10 kcal mol\textsuperscript{-1} \SI{}{\angstrom}\,\textsuperscript{-2}) on graphene atoms (no restraint on dendrimer). The Langevin thermostat and Berendsen barostat were used to control the temperature at 300 K and pressure at 1 bar, respectively, with a collision frequency of 1 ps$^{-1}$ and pressure coupling constant of 0.5 ps. Finally, we performed 50 ns long NVT production dynamics at 300 K, using a Langevin thermostat. The electrostatic interactions were evaluated using the particle mesh Ewald (PME) method ~\cite{darden1993particle} with a real space cut off of \SI{9}{\angstrom}  and a reciprocal space convergence tolerance of $5 \times 10^{-4}$. The SHAKE algorithm ~\cite{ryckaert1977numerical} with a tolerance of \SI{0.005}{\angstrom} was used to constrain all bonds involving hydrogen atoms, as well as the H-O-H angle in water, to their equilibrium positions. This allowed us to use larger integration time steps of 2 fs. The trajectory from the final 10 ns of the MD simulations was used for statistical data analysis.  
    
   
    We obtained the electrostatic potential across the cell by solving the 2D Poisson's equation with $xy$-symmetry. 
    We set the field to be zero at both the ends of the box (i.e., Neumann BC) \cite{boda2013calculating}, and  set the potential to be zero at the middle of the simulation cell. With these two BCs, one can formulate the expression for the potential as,
   
    \beq
    \label{eq:ccm_pot}
    \psi(z) = - \frac{1}{{\epsilon}_0} \int_{-D}^{z} dz^{\prime} \int_{-D}^{z^{\prime}} \rho(z^{\prime \prime}) dz^{\prime \prime} - \psi(0).
    \eeq
    where, $\rho(z)$ is the charge density along the sheet's normal and $\psi(0)$ is the potential at the mid-point of the 
    simulation box. The positive and negative electrodes are located at $-D$ and $+D$, respectively. Once the potential profiles were obtained, one can define the capacitance as:~\cite{park2016computer}

  \beq
    C_s = \frac{\lvert {\sigma}_s \rvert}{\Delta \Delta \psi_{cell}}, \quad C_M = C_s \frac{A_0}{2M}.
    \label{eq:cap}
 \eeq
    Here, 
  \beq
    \Delta \Delta \psi_{cell} = \Delta \psi_{cell}(charged) - \Delta \psi_{cell}(discharged)
    \label{eq:vcell}
  \eeq
    and,
    $$\Delta \psi_{cell} = \psi(z = -D) - \psi(z = +D)$$ 
    where $\sigma_s$, $A_0$ and $M$ are the charge density, area 
    and mass of the electrode, respectively, 
    $C_s$ is the area-specific capacitance with units of \SI{}{\micro\farad\per\square\centi\meter} and 
    $C_M$ is the mass-specific capacitance with units of \SI{}{\farad\per\gram} ~\cite{vatamanu2015non}.

\subsection{Constant Potential Method (CPM)}
 The CCM approach neglects the charge fluctuations on the electrode induced by local density fluctuations in the electrolyte solution.
To explicitly takes into account such physics, we also considered the CPM method which allows the charges on the electrode atoms to vary while maintaining
a constant potential. We follow the implementation of the CPM method developed by Reed \textit{et al.} and later further improved by Gingrich \textit{et al.} ~\cite{gingrich2010ewald}, based on earlier
work of Siepmann  \textit{et al.} ~\cite{siepmann1998influence}.
Here, the electric potential ($\psi_i$) on each electrode atom is constrained at each simulation step to be equal to a preset applied external potential $V_{ext}$, which is held constant. 
The charge on each electrode atom $q_i$(where i is the atom index of the electrode), is calculated self-consistently to satisfy the constraint condition:~\cite{reed2007electrochemical}

\begin{equation}
    \psi_{i}=\frac{\partial U}{\partial q_i}=V_{ext}
\end{equation}
where $U$ is the total coulomb energy of the system.

The starting structures for the CPM simulations were the equilibrated structures of the CCM simulations (discharge cases).
All CPM simulations were performed using the LAMMPS \cite{wang2014evaluation} MD package. \SI{20}{\nano\second} NVT production simulations were performed using Nos\'{e}-Hoover thermostat at \SI{300}{\kelvin} with a relaxation time of \SI{100}{\femto\second}
, which was long enough to achieve the total charge equilibration of the electrodes. 
The cutoff used for the non-bonded interactions was \SI{9}{\angstrom}. The long-range electrostatic interactions were calculated by the particle-particle-particle-mesh Ewald summation method with a tolerance of $10^{-4}$. Instead of using 2d-periodic Ewald
sums, we have used 3d-periodic Ewald sums with shape corrections  in this work to enhance  the calculation speed, with a volume factor set to 3 \cite{yeh1999ewald}. Parameter of Gaussian charge distribution for electrode atom was \SI{19.79}{\per\nano\meter} \cite {reed2007electrochemical}. All the result are reported here are the last \SI{10}{\nano\second} statistical averages.

       The electrodes are maintained at a constant potential  $ \pm V_{ext} $ using the 
       constant potential method (conp) module in LAMMPS ~\cite{wang2014evaluation}. 
       To calculate the differential capacitance, we first calculated the averaged equilibrated charge density of the electrodes (which is a direct output
       from the CPM simulations) and applied the following equation: 
       \begin{equation}
           C_{diff} = \frac{\, d \langle\sigma\rangle}{\, d V_{ext}}
       \end{equation}
       where, $\langle\sigma\rangle $ is the equilibrated average surface charge density on the electrode. The mass-specific capacitance can be calculated from Eq. \ref{eq:cap}.
     
Note that, one can employ constant pH method \cite{baptista2002constant, machuqueiro2006constant} also to simulate dendrimer and dendrimer coated graphene electrodes  to investigate their behaviour. However, constant pH method is best suited for systems where protonation state of the system is not very well known and has been used for some protein simulations. The situation is different in case of dendrimer where titration experiments give the protonation state of dendrimer as a function of pH very accurately \cite{van1998acid}. 
That is why in the literature dendrimer simulations are done taking into account the protonation states corresponding to a pH as has been done in the present study. 
It is worth mentioning here that constant pH simulation for dendrimer also \cite{reis2018role} has shown to produce similar results like pH dependent swelling using the same protocol used in our work.
But constant pH calculations take significantly large computational cost.

\section{\label{sec:level3} Results and Discussions}
\subsection{ CCM method }
\subsubsection{EDL structure}
\begin{figure}[ht!]
    \centering
    \includegraphics[width=\linewidth]{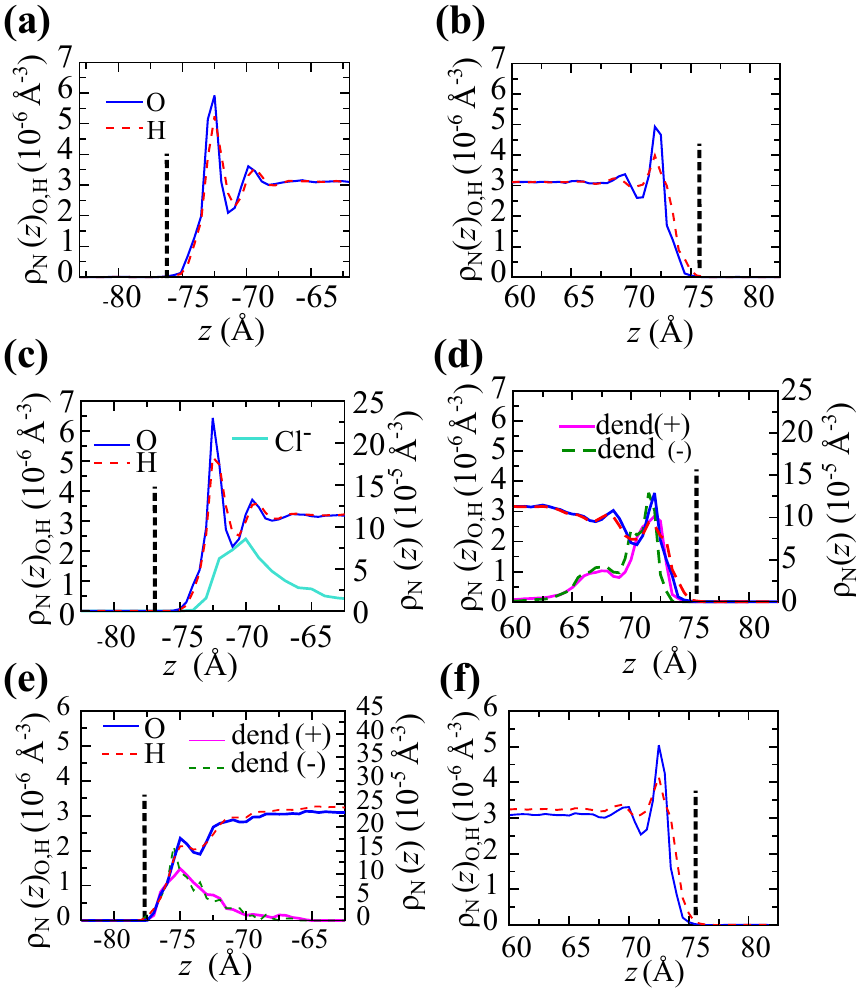}
    \caption{Number density profiles for BAREGP at the (a) left and (b) right electrodes, obtained using the CCM method. (c), (d) and (e), (f) are the profiles for G2PEL and GPDEN cases, at the left, right electrodes, respectively. The density profile
    of any given species is
    normalized by its total number in every case. We denote the positive and negative electrodes
    as $L$ and $R$, denoting the left and right electrodes of the system. Number densities for Hydrogen and Oxygen scale with left y-axis whereas for dendrimer and $Cl^{-}$ ions, it scales with right y-axis. Black dashed lines schematically represent the position of the graphene sheet.}
    \label{fig:numbden}
\end{figure}

The structure of the electrode/electrolyte interface is an important property that gauges the
performance of a supercapacitor. Hence, to understand the nature of the EDL structure and the distribution of the positive and the negatively charged entities of the system, we calculated their number densities across the cell. In Fig. \ref{fig:numbden} we plot the
normalized density profiles ($\rho_N(z)$) of the relevant charged species of the system, 
defined as, 
\begin{equation}
\rho_N(z)_{\text{species}} = \frac{1}{N_{\text{species}}}\frac{n(z,z+dz)}{L_x\, L_y\, dz},     
\end{equation}
where $N_{\text{species}}$ is the total number of a given species in the system, and $n(z,z+dz)$ is the number of 
the atoms of the species located between $z$ and $z+dz$. $L_x$ and $L_y$ are the box dimensions in the
$x$ and $y$ directions, respectively. By definition $\int_{-D}^{D} \rho_N(z) L_x\, L_y\, dz =1.$

In the case of BAREGP and GPDEN, there are no mobile anions or cations that would adsorb at the 
electrode's interface, thus electrostatic screening of the charged interface is only achieved by reorientation of the dipole of the water molecules. This fact, coupled with the nature of the electrode charge distribution leads to notable changes in the interfacial water structure. For BAREGP, we find in \ref{fig:numbden} (a) and
(b), significant densification of the 1st layer waters compared to the bulk. While density fluctuations are commonly observed in fluids next to a hard wall (due to the reduction in dimensionality) \cite{allen2017computer}, we find that the 1st shell water density and layer thickness on the left (positively charged) BAREGP electrode is larger than on the right (negatively charged), reflecting the directionality of the hydrogen bonding in water, which competes with the electrostatic screening effect that requires the water dipole point into the negative electrode (i.e., both hydrogens closer to the surface). The net effect is more disordered 1st layer at the negative electrode.

Turning our attention to the systems with the dendrimer, we find that
the 1st layer water structure at the positively charged electrode of GPDEN (Fig. \ref{fig:numbden} (e)) differs significantly  
from that of BAREGP and G2PEL (Fig. \ref{fig:numbden} (a)). This reflects the inhomogenous charge distribution in GPDEN, where we have the dendrimer coated graphene as the positively charged
electrode (Fig. \ref{fig:numbden} (e)). In GPDEN, the water molecules can penetrate
deeper in to the electrode (i.e., dendrimer), so that the effective surface charge of the positively charged electrode is well compensated. By contrast, in the G2PEL system, the adsorption of the positively charged dendrimer at the
right electrode (negatively charged) leads to best possible screening of the electrode's charge.
(Fig. \ref{fig:numbden}(d)). Similarly, the mobile Cl\textsuperscript{-} anions compensate the positive charge of the left electrode (Fig. \ref{fig:numbden} (c)). Overall these microscopic EDL reorganization effects have a significant impact on the cell capacitance, as will be discussed in section III-B.

   \begin{figure}[h!]
    \begin{center}
      \includegraphics[width=\linewidth]{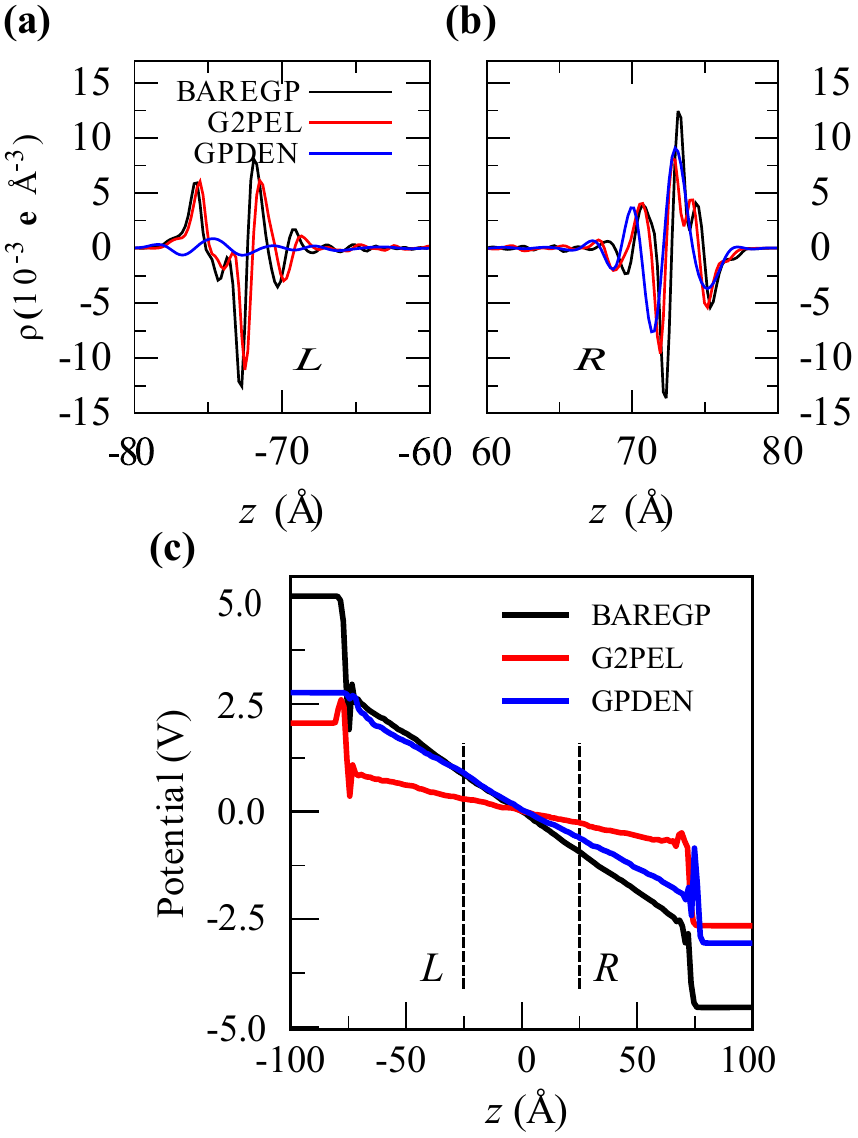}
  \caption{(a),(b) In-plane averaged equilibrium charge density along the $z$ direction. subscripts $L$ and $R$ designate left and right respectively. (c) Potential profile across the cell for BAREGP, G2PEL and GPDEN cases for CCM case. Presence of the dendrimer as an electrolyte and electrode reduces the potential difference across the simulation cell.}
      \label{fig:potential}
    \end{center}
  \end{figure}
\subsubsection{Potential profiles and capacitance values}
The total charge densities of the systems at the left and the right electrodes are presented in
Fig. \ref{fig:potential} (a) and (b), respectively, and encapsulates the results of the EDL structure in section 
III-A. For example, at the left electrode, the charge density for GPDEN is significantly
less than BAREGP (Fig. \ref{fig:potential} (a)), due to the better
penetration of the water molecules in to the electrode in the latter case.
Similarly, we find that the charge density of G2PEL at the right electrode is less than that of BAREGP and GPDEN (Fig. \ref{fig:potential} (b)) due to the better screening of the electrode's charge by dendrimer adsorption (Fig. \ref{fig:numbden} (d)). 
Further, we obtained the electrostatic potential ($\psi(z)$) profiles across the cell using Eq. \ref{eq:ccm_pot} as illustrated in Fig. \ref{fig:potential} (c) using the charge density data. We found a smaller potential difference across the electrodes for G2PEL and GPDEN compared to that observed in BAREGP (Table ~\ref{tb:ccm}). This results directly from the increased screening of the surface charge by the dendrimer, resulting in a reduced electric field and hence a
smaller potential in the bulk region of the cell for G2PEL and GPDEN. Furthermore, in BAREGP and GPDEN, the large, finite slope in the potential profiles in the bulk region suggests a strong electric field in the middle of the simulation box, which is due to the heavily charged electrodes and the absence of any potential screening effect due to free ionic salts, as evidenced by the reduced slope in the G2PEL system. Nevertheless, we do find a modulation of the potential drop at the electrode upon introduction of the dendrimer in the GPDEN system that is largely absent in BAREGP, which suggests better electrostatic screening compare to pristine graphene as an electrode.

\begin{table}[!h]
    \centering
    \caption{The potential difference across the cell and the capacitance values calculated using CCM model.}
   \begin{ruledtabular} 
       \begin{tabular}{  c  c  c  c c}     
           System 
           & $ \Delta \psi_{cell} (dis)(V)$
           & $ \Delta \psi_{cell}  (ch) (V)$ 
           & $ C_{s} \left (\frac{\mu F}{cm^2} \right ) $ 
           & $ C_{M}\left(\frac{F }{g}\right)$ \\ [0.5ex]
           \hline
           BAREGP & 0.12 & 9.53  & 1.18 & 6.58 \\ [0.5ex]
                    GPDEN &   0.12 & 5.86 &  1.95 & 11.03 \\ [0.5ex]
                    G2PEL & 0.04 & 4.75 & 2.35 & 13.14 \\ [0.5ex]
       \end{tabular}
       \label{tb:ccm}
   \end{ruledtabular} 
     \end{table} 
To test this hypothesis and quantify the effect, we calculated the cell capacitance using Eq. \eqref{eq:cap} for all three systems and tabulated the values in Table ~\ref{tb:ccm}. This required us to perform both \emph{charged} and \emph{discharged} simulations in order compute the potential difference across the cell (Eq. \ref{eq:vcell}). While we performed these two 
sets of simulations for BAREGP and G2PEL, for GPDEN case we performed only \emph{charged}
simulations, since we considered the discharged GPDEN case to be the same as the BAREGP system. 

We found significantly higher capacitance in the two systems with the dendrimer, compared to BAREGP, with an increase of  $65.25 \%$ and $99.15 \%$ for the dendrimer in electrode (GPDEN) and electrolyte (G2PEL) systems, respectively. 
To further understand the molecular origins of this effect, we divided our simulation cells into left ($z=\SI{-100}{\angstrom}$ to $z= \SI{-25}{\angstrom}$) and right ($z=\SI{25}{\angstrom}$  to  $z= \SI{100}{\angstrom}$) sub-boxes.
We consider a virtual parallel plate capacitor whose electrodes are now represented by these left and right regions, indicated by the two parallel line separators in Fig. \ref{fig:potential}.

\begin{table}[!h]
    \centering
      \caption{Surface charge density of the virtual parallel plate capacitor. All the cases correspond
    	to the \emph{charged} simulation set.}
   \begin{tabular}{ p{2.5cm}  p{2.75cm}  p{2.75cm}  }
        \hline
        \hline
        System & $\sigma_L \left (10^{-4}e \SI{}{\angstrom}^{-2} \right )$ & $\sigma_R\left (10^{-4}e\SI{}{\angstrom}^{-2} \right )$\\ [0.5ex]
        \hline
         BAREGP & 1.98 $\pm $0.32 & -2.08 $\pm$ 0.20 \\ [0.5ex]
         G2PEL & 0.58 $\pm $ 0.28 & -0.63 $\pm$ 0.19 \\ [0.5ex]
         GPDEN & 1.87 $\pm$ 0.30 & -1.54 $\pm$ 0.31\\ [0.5ex]
         \hline
         \hline
    \end{tabular}
  
    \label{tab:ch_dens}
\end{table}
The surface charge density $\sigma_L$ was then calculated by integrating the charge density across each sub-box, as below: 
\begin{align}
    \sigma_{L}(z) &= \int_{-D}^{z} \rho (z) dz, \quad z \in \left[ -55, -25 \right] \\ 
    \sigma_{R}(z) &= \int_{z}^{D} \rho (z) dz, \quad \quad z \in \left[ 25, 55 \right] 
\end{align}
This allows us to reduce the fluctuations in a given bin during the simulation.
The average values for $\sigma$ are obtained using, $$\sigma_{L,R} = \langle \sigma_{L,R} (z)\rangle$$ and are presented in the Table \ref{tab:ch_dens}.

For both the G2PEL and GPDEN cases, the net surface charge on the virtual parallel plate capacitor was found to be smaller than in BAREGP, again suggesting that the electrode's charge is better compensated in the 
presence of the dendrimer. Here, in G2PEL, the solvated G2 PAMAM dendrimer behaves as a dielectric screening medium that decreases the strength of electric field, thus decreasing the potential and increasing the capacitance. Indeed, the dendrimer in GPDEN can be thought of as a 
porous electrode, where the availability of
the charged groups accessible to the water molecules gives
rise to the lowered potential difference. 

\subsection{Capacitance values: CPM method}

\begin{figure}[!h]
    \begin{center}
    \includegraphics[width=\linewidth]{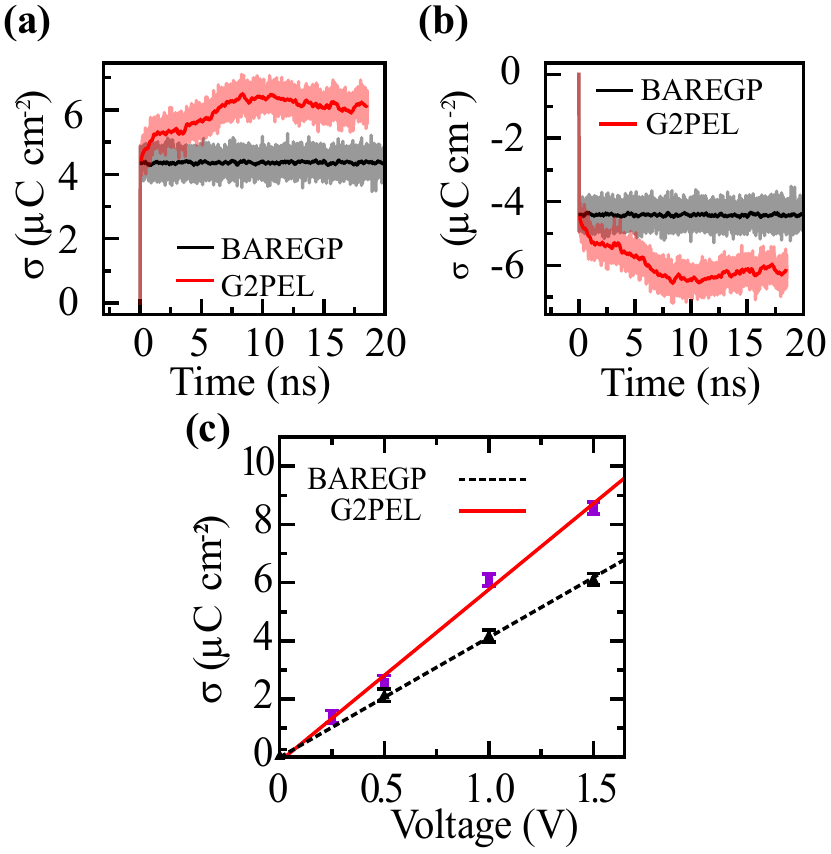}
     \caption{ Time evolution of the charge per unit area deposited on the (a) positive and (b) negative electrodes for BAREGP and G2PEL cases. The electrodes are maintained at fixed constant potentials $\pm 1 V $, using CPM algorithm. Presence of dendrimers accumulates more charge at electrodes to maintain $ 1 V $.
     (c) Surface charge density on the positive electrode as a function of applied potential across the cell. The lines represent linear fits of the data.  The corresponding slopes give the capacitance \SI{5.84}{\micro\farad\per\square\centi\meter} and   \SI{4.18}{\micro\farad\per\square\centi\meter} for G2PEL and BAREGP cases, respectively.}   
    \label{fig:charge_cpm}
    \end{center}
\end{figure}
Fig. \ref{fig:charge_cpm}(a),(b) show the accumulation of positive and negative charges density on the two electrodes in order to maintain $\pm 1V $, obtained using the CPM method. Equilibrium surface charge densities  for all the voltage are included in Fig. \ref{fig:all_volt} in Appendix.

Convergence was typically observed after $\sim 10$ ns of MD and we found that the presence of protonated G2 PAMAM dendrimer significantly affected the charge distributions, compared to BAREGP.
In Fig. \ref{fig:charge_cpm}(c) we depict the charge density $\sigma $ of the positive electrode as a function of applied voltage for the BAREGP and G2PEL systems
(the negative electrode acquired an opposite charge density).
In both the cases, we found that the surface charge density of the electrode scaled linearly with applied potential, which implies a constant value of the differential capacitance over the potential range.

\begin{table}[!ht]
    \centering
    \caption{Differential capacitance values calculated using CPM model.}
     \begin{ruledtabular} 
       \begin{tabular}{ c c  c  } 
           
           System
           & $ C_{s} \left( \SI{}{\micro\farad\per\square\centi\meter}\right) $ 
           & $ C_{M}\left (\SI{}{\farad\per\gram}\right )$ \\ [0.5ex]
           \hline
           BAREGP   & 4.18 & 23.35 \\ [0.5ex]
           G2PEL  & 5.84 & 32.62 \\ [0.5ex]
 
    \end{tabular}
    \label{tb:cpm}
     \end{ruledtabular} 
\end{table} 
The values of the capacitance were then computed through a linear regression of the charge density data, shown in Fig. \ref{fig:charge_cpm}c, with the results being \SI{4.18}{\micro\farad\per\square\centi\meter} and   \SI{5.84}{\micro\farad\per\square\centi\meter} for BAREGP and G2PEL cases respectively (Table ~\ref{tb:cpm}). These capacitance and the surface charge density data are consistent with those reported in literature for BAREGP case. ~\cite{jeanmairet2019study}
Notably, we found a $39\%$ increase in capacitance in G2PEL, compared to BAREGP.

\begin{figure}[!h]
    \centering
    \includegraphics[width=\linewidth]{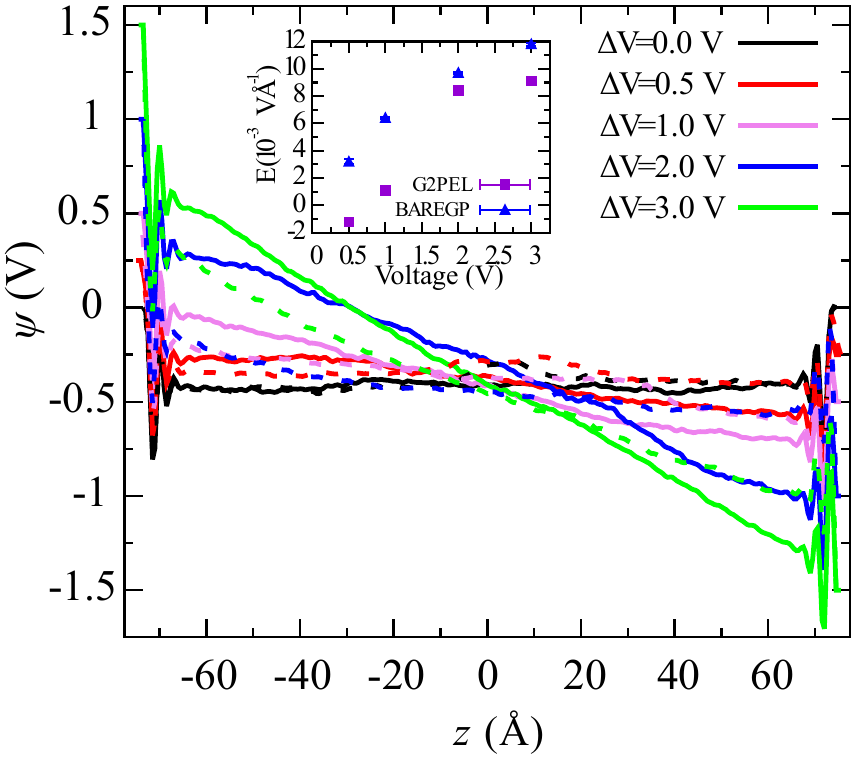}
    \caption{Poisson potential ($\psi$) across the cell computed using Eq. \eqref{eq:psi_z} for different values of the applied voltage $ \Delta V $. Different colors represent different voltage. Solid line and dash represent BAREGP and G2PEL case respectively. The inset of Fig. depicts electric field inside the bulk region as a function of the applied potential difference in a cell.} 
    \label{fig:poission_pot}
\end{figure}

Further insights into the origin of the increased capacitance was obtained by computing the potential profile of the cell using the Poisson equation \cite{boda2013calculating}:
\begin{equation}\label{eq:psi_z}
    \Delta \psi(z)=  
    - \frac{1}{{\epsilon}_0} \int_{-L}^{z} dz^{\prime} \int_{-L}^{z^{\prime}} \rho(z^{\prime \prime}) dz^{\prime \prime} + C_1(x-L) +V_{ext},
\end{equation}
\beq
  C_1=  
    \frac{1}{L_2- L_1}\left[\Delta V_{ext} + \frac{1}{{\epsilon}_0} \int_{-L}^{L} dz^{\prime} \int_{-L}^{z^{\prime}} \rho(z^{\prime \prime}) dz^{\prime \prime} \right].
\eeq
Here we use Dirichlet BC,
    $\psi (L_1=-D)= V $  and $\psi (L_2=D) = -V $.
  Fig. \ref{fig:poission_pot} shows that the potential maintained a constant value within the electrode, oscillates  in a region of \SI{15}{\angstrom} near the electrode interface, before exhibiting a linear profile in the bulk region.  
  The oscillation of the potential close to the electrode is due to water layering, resulting from intrinsic entropy-dominated forces at low voltage and stronger graphene-water interactions and electrostatic binding at larger potentials \cite{pascal2021entropic}.

\begin{table}[!h]
    \centering
    \caption{Electric field  in bulk region of electrolyte of BAREGP and G2PEL at different voltage}
  \begin{ruledtabular}   

    \begin{tabular}{  c  c  c  c }     
$\Delta $V (V) & E($10^{-3} V $ \SI{}{\angstrom}$^{-1})$  &  E($10^{-3} V $ \SI{}{\angstrom}$^{-1})$  \\ [0.5ex]
          &  in BAREGP &   in G2PEL \\ [0.5ex]
  \hline
 0.5 & 3.2 $\pm$ 0.2 & 1.2 $\pm$ 0.1 \\ [0.5ex]
 1.0 & 6.4 $\pm$ 0.1 & 1.1 $\pm$ 0.2  \\ [0.5ex]
 2.0 & 9.7 $\pm$ 0.1 &  8.4 $\pm$ 0.1 \\ [0.5ex]
 3.0 & 11.8 $\pm$ 0.1 & 9.1 $\pm$ 0.2  \\ [0.5ex]

\end{tabular}
\label{tb:ef}
 \end{ruledtabular} 
\end{table} 
  To understand how dendrimer affects the potential profile we calculated the electric field inside the bulk region (Table \ref{tb:ef}). Here, we used the negative gradient of the potential in the bulk, from $ - L^{\prime} $ to  $ L^{\prime} $, where $L^{\prime} = \frac{D}{2.5}$ $\rightarrow$ $\frac{D}{3.3}$ . 
We found that as the voltage increases, so does the electric field in the bulk region (Table \ref{tb:ef}). Fig. \ref{fig:poission_pot} shows that for G2PEL, the strength of the electric field was reduced in the bulk region, as compared to BAREGP, due to electrostatic screening of the dendrimer at all voltages. In other words, in order to maintain the required voltage, the G2PEL system has to store more charge than BAREGP.
\subsection{Comparison between CCM and CPM approaches}

To compare the CCM and CPM method, we performed additional simulations using the CCM method, where total charge on each electrode was taken as the average equilibrated charge in the CPM method at  $\pm 1 V $, distributed evenly on each carbon atom of the graphene sheet.

\begin{figure}[!ht]
    \centering
    \includegraphics[width=0.9\linewidth]{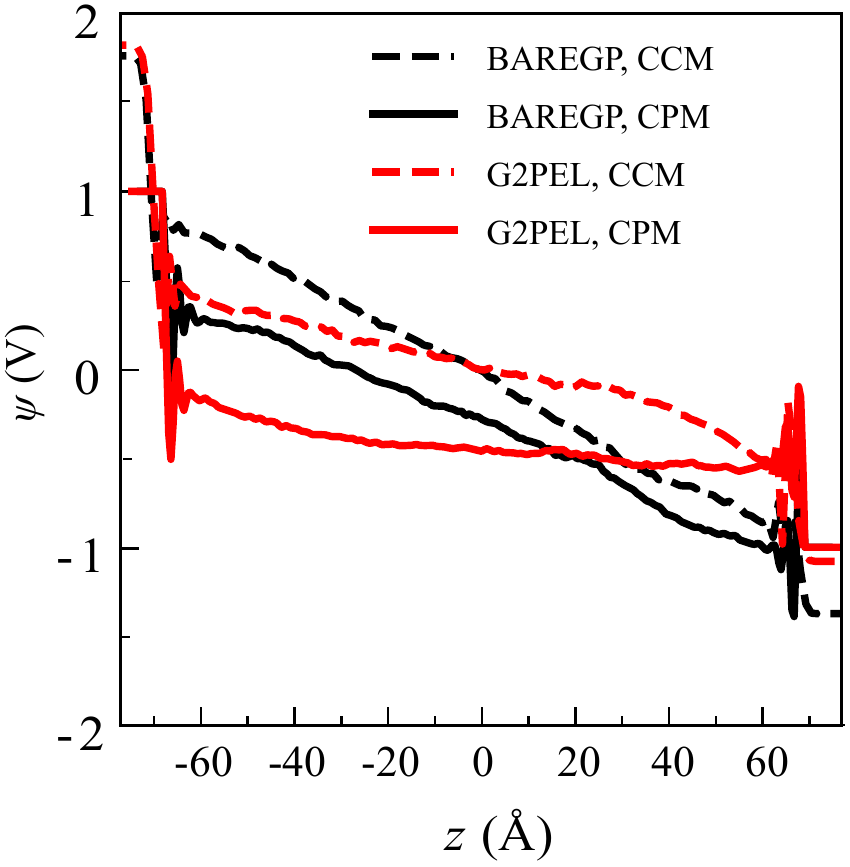}
    \caption{Poisson's potential $(\psi)$ profile for constant charge simulations, where a fixed charge is put on every atom of the electrodes corresponding to the average equilibrated charge on the electrode correspond to $\pm 1V$. Left: positive electrode; right: negative electrode}
    \label{fig:CCM-1V}
\end{figure}

\begin{table}[!h]
    \centering
     \caption{Comparison of capacitance computed using CPM and CCM respectively.}
   \begin{ruledtabular} 
\begin{tabular}{ c  c    c c    c }
\multirow{2}{*}{System} & \multicolumn{2}{c}{$\Delta \psi (V)$} & %
    \multicolumn{2}{c}{$ C_{s} \left(\SI{}{\micro\farad\per\square\centi\meter}\right)$  }\\ [1.5ex]

 & CPM & CCM & CPM & CCM \\ [1.5ex]
\hline

 BAREGP& 2.0 & 3.14 & 2.08 & 1.32 \\ [0.5ex]
 G2PEL& 2.0 & 2.85 & 2.59 & 1.89 \\ [0.5ex]

\end{tabular}
\end{ruledtabular}

    \label{tab:ccm_cpm} 

\end{table}

Fig. \ref{fig:CCM-1V} reports the Poisson's potential across the simulation cell for both methods.
The solid curves are the results obtained from CPM method using Dirichlet BCs (\ref{eq:psi_z}) and dashed curved are the potential profile correspond to the CCM method using Newman's BCs (\ref{eq:ccm_pot}).
The corresponding data is shown in Table \ref{tab:ccm_cpm}. Our results using the CPM method compares well to reported values in the literature ~\cite{jeanmairet2019study}.
We found that the although the CCM method predicted consistant capacitive enhancements in the G2PEL system compared to BAREGP, it generally overestimated the value of the potential in both cases. Thus, we conclude that while CCM method may be adequate to obtain  qualitatively insights into the electrochemical behaviour of these systems, more detailed, quantitative results will require the CPM method.

\section{\label{sec:level4} Conclusions and Future outlook}
To summarize, we have used extensive MD simulations to show that the presence of a protonated G2 PAMAM dendrimer, in aqueous solutions or grafted on the electrode, enhances the capacitance and reduces the potential across the cell significantly compared to pristine graphene electrodes.
The EDL structure and overall cell performance was shown to be a dramatic function of the dendrimer morphology (free or grafted) and chemistry (charged or neutral), suggesting two different ways to utilize dendrimers with the aim of improving the capacitances
of the graphene-based aqueous supercapacitors. 
Further, these results advance our understanding of the role of macro-molecules in modifying the properties of electrochemical systems and suggest important avenues for predictive design for improved efficiencies.

\begin{acknowledgments}
TM thanks MoE, India, for financial support in the form of scholarship. We also acknowledge computational support through TUE-CMS, IISc funded by DST, India. We also thank Supercomputer Education and Research Center (SERC), IISc, for providing supercomputer time at CRAY, SAHASRAT machine. PKM acknowledges funding through SERB, IRHPA (No.IPA/2020/000034).
\end{acknowledgments}
\appendix
\section{Charge density profiles as a function of the applied voltage}

  \begin{figure}[h!]
    \begin{center}
\includegraphics[width=\linewidth]{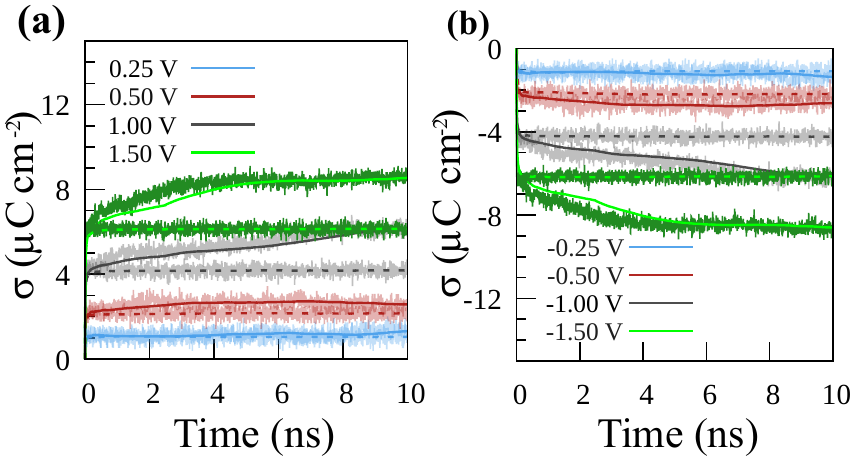}
\caption{(a),(b) equilibrium charge densities in positive electrode and negative electrode respectively. Solid and dashed line represent G2PEL and BAREGP case respectively. Different color represent different voltage. For all the voltage G2PEL case required more charge on electrode to maintain a fixed voltage compared to BAREGP.}
\label{fig:all_volt}
 \end{center}
 \end{figure}

\bibliography{ref}


\end{document}